# From Fragments to an Integrated European Holocaust Research Infrastructure

*Tobias Blanke, Veerle Vanden Daelen, Michal Frankl, Conny Kristel, Kepa J. Rodriguez & Reto Speck*

In order to keep remembering why it was judged so important to build a new Europe 'out of the crematoria of Auschwitz', the 'vital link' between Europe's past and Europe's present should, according to the British historian Tony Judt, be taught over and over again (cf. Judt 2005: 830 f.). Every generation of European historians should re-interpret the message of the Second World War and of the Holocaust anew. To be able to discuss, negotiate and teach Europe's past, the historical research needs to become truly transnational and transcend national borders. In order to achieve this and to enable new, innovative forms of Holocaust research, the European Holocaust Research Infrastructure (EHRI) has been set up by the European Union to create a sustainable complex of services for researchers. EHRI will bring together information about dispersed collections, based on currently more than 20 partner organisations in 13 countries and many other archives. EHRI, which brings together historians, archivists and specialists in digital humanities, strives to develop innovative on-line tools for finding, researching and sharing knowledge about the Holocaust. While connecting information about Holocaust collections, it strives to create tools and approaches applicable to other digital archival projects.

Trans-national research into the Holocaust can be a very challenging undertaking. Holocaust studies rely more than other fields of research on a huge variety of archives. Holocaust archives are fragmented and dispersed all over the world, making access complicated, if not impossible and very time-consuming. Researchers therefore have to deal with different archival systems, laws, rules on access and copyright, cataloguing systems, or in short: different archival cultures.

The fragmentation of archival sources does not only result from the wide geographic scope of the Holocaust, but also from the Nazi attempts to destroy the evidence. Moreover, and closely related to the geographical scope, Holocaust sources were written in many different languages, the language of the documents not necessarily being the same as the cataloguing and descrip-tion language, thus further complicating the picture. Furthermore, Holocaust survivors migrated to places across the world taking documentation with them, and a plethora of documentation projects was developed after the Second World War.

Next to these challenges some new ones have emerged recently. In the past decades more and more specific collections have been set up, in many regional centres on research and commemoration. The opening up of archives in Eastern Europe, and in particular in Eastern Germany, and the opening of formerly classified archives in Western Europe has resulted in a substantial increase in available source material. At the same time, the number of institutions in European countries that hold Holocaust-related collections and are active in the field of research and commemoration, has increased since 1989 (especially in Eastern Europe but also in Germany and most other European states). These institutions, old and new, have their own collections and their own (increasingly digital) archival infrastructures, which often do not support scholarly requirements. Different institutions use their own distinct systems and different metadata schemas. Many different languages are used in the original documents as well as in catalogues, necessitating translation and hampering comparability.

Until 1989 the United States, Israel and Western Europe were the main centers for Holocaust research. Auschwitz became the symbol for the Holocaust worldwide, because it was the largest death camp, but also because it was the camp where Jews from Western and Central Europe were murdered. However, the vast majority of Holocaust victims lived and was murdered in Eastern Europe. Research and documentation on this part of Europe is still far more difficult to conduct than on Germany or Western Europe.

Finally, one of the major challenges for every scholar of the Holocaust is to deal appropriately with the prevalence of perpetrators' sources to avoid muting the voices of the persecuted Jews. The documents of Jewish organizations or relief organizations often followed the fate of their owners; they were in many cases destroyed or dispersed. For instance, to gain insight into the activity of the Jewish refugee organisations in Prague in the 1930s, a researcher has to study the fragments of reports saved in several archives, especially in the USA, Israel, Czech Republic and Germany.[1] And while there are numerous testimonies given by Holocaust survivors after the liberation, original diaries, letters and/or testimonies from the time of persecution are more difficult to find. Over the last years a growing consensus has emerged in Holocaust historiography that Jewish sources and views have to be more integrated into the narrative(s) of the Holocaust.[2]

---

1 See the information about sources in Čapková/Frankl (2012: 17 f.).

By connecting collections from different archives and countries with research, EHRI can make a significant contribution towards more integrated research on the Holocaust. This article will discuss our current efforts and present initial results in some key areas midway through the project. First we will present some of our innovations such as our identification work to discover relevant collections and archives and bringing their descriptions together into a single point of access: the EHRI portal. We will then dig deeper into the problem of the dispersion of Holocaust material by discussing a particular case study: the Terezín (Theresienstadt) ghetto's documentation, and the digital research guide that we are developing to address the specific challenges of the Terezín use case.

## Background and State of the Art

EHRI did not have to start from scratch: organisations throughout the world have already done excellent work in collecting and saving documents, objects, photos, film and art related to the Holocaust. Therefore, on the most basic level, EHRI started by identifying and merging information on institutions which hold Holocaust-related archival material, the collection-holding institutions (CHIs). The EHRI identification work's starting points were the *Directory of Holocaust-related Archives* of the Conference on Jewish Material Claims Against Germany;[3] the *Guide des archives sur la Shoah* of the Mémorial de la Shoah;[4] the list of institutions with which Yad Vashem has worked and has copied material from.[5] This compiled list was further completed with information from the United States Holocaust Memorial

---

[2] See for instance Andrea Löw's recent book about the ghetto in Lodź (Löw 2006). More recently see Laura Jockusch's book about the early Jewish documentation projects (Jockusch 2012).

[3] http://www.claimscon.org/archivist_forum/archive_search.asp

[4] http://www.memorialdelashoah.org/b_content/getContentFromTopNavAction.do?navId=12

[5] As a Consortium partner, Yad Vashem provided EHRI with internal overviews of institutions surveyed (Yad Vashem, internal documents).

Museum[6] and aggregators such as national research portals on archival sources on the Second World War (e.g. http://www.archievenwo2.nl/, or http://www.ns-quellen.at/), national archival guides, experts and published studies on the subject.[7] All the collected information on CHIs has been entered into a central repository. This repository uses the ICA-AtoM software, which is fully open source and complies with the ISDIAH standard.[8]

*Figure 1* Screenshot of the current EHRI database (work version)
(source: http://icaatom.ehri-project.eu/index.php )

---

6 mostly via the "Archival Guide to the Collections" (which provide a general overview of the collections of textual records available in the Museum's Archives), and via the "Archival Finding Aids" (detailed inventories and finding aids that have been produced for selected collections in the Museum's Archives) (www.ushmm.org)

7 An overview of the used archival guides will be published onto the EHRI portal in the course of 2013.

8 For general information ICA-AtoM we refer to https://www.ica-atom.org/. The EHRI version of ICA-AtoM has the following address: http://icaatom.ehri-project.eu/.

The next step of our general workflow is the identification of collections within the CHIs, which hold Holocaust-relevant material. All EHRI collection descriptions follow the ISAD (G) standard and will be – as much as possible – provided in English. The descriptions of these collections can be brought into the portal, and linked to the CHIs' information via two major pathways: harvesting or manual data entry. Using harvesting, the project would have the highest level of sustainability. However, in the first instance, identification will also happen manually, as many partner institutions currently do not provide a harvesting service. To have as many institutions as possible join the 'digital connection' group, a tool and example are being developed to help institutions to meet the future requirements of EHRI if they can and want to. Our surveying work, which we describe next, collects all this technical information next to its main focus on identifying sources.

In order to provide an overview that is as complete as possible, all sources related to the Holocaust could and should in principle find their place on the EHRI portal. EHRI does not want to exclude, but for pragmatic reasons some criteria of prioritization have to be put in place. For instance, EHRI has decided to prioritize the identification of victims' sources, because perpetrator sources are more numerous and less dispersed and therefore better known. Special attention is given to identifying Eastern (and Central) Europe sources, as these are the places where most victims perished, and where sources – compared to the West – have been less inventoried and made accessible.

EHRI's surveying work starts with Germany and its allies, and the countries occupied by the Axis Alliance, but EHRI will not stop here and further countries will get included in the further stages of the project, too. Any place where Holocaust-related sources can be found should in the end be included. In order to structure the identification work, EHRI developed a concise internal working definition of the Holocaust, which we see as a tool providing the EHRI consortium with a framework.

To complete EHRI's identification work within a systematic and structured framework, national reports are being written, which will give a short overview on a country's Holocaust history and its archives. All EHRI national reports follow the same general structure. First, in two short paragraphs a general overview of a country's history during the Second World War is given. The first paragraph covers questions of statehood as well as German rule and influence, while the second paragraph focuses on Holocaust history and also includes information on the size of the pre-war Jewish community as compared to the total population of a country and an estimated number of Jewish victims. In a second section the reports describe briefly the archival situation. Again, the same structure is used for all national reports. A first paragraph deals with the archival culture of a country and how its archives are organized (centralized system, role of the state, and legislation can be addressed here). The second paragraph gives more information on which archives are most relevant for Holocaust research. The third part of the reports will give a concise overview of EHRI's research in the country. This results in a one to (at most) two pages executive summary per country. In consulting these summaries, the reader should get a clear and concise overview of why the country is being researched and what the current state of knowledge and access to Holocaust-related archives in this country is, including a concise overview of EHRI's identification efforts.

In doing all this, EHRI supports new research and enables historiographical progress by providing access and by sharing information on sources relatively unknown until now. EHRI's identification work assists the Holocaust studies' globetrotters so that they know as much as possible before they start planning travels. Researchers will be able to evaluate more easily which sources are located where, whether there are duplicates available in other repositories, etc.

Next to this more global overview of resources in national reports and collection descriptions, Holocaust studies is also supported by specific research guides giving detailed access to individual topics such as the Terezín ghetto.

## Dispersion of Records and Challenges for Holocaust Research – The Terezín Case

The ghetto in Terezín (Theresienstadt in German) was one of the major places of suffering and death of Jews from Bohemia and Moravia, Germany, Austria, the Netherlands, Denmark and other European countries. Out of approx. 150,000 prisoners, over 30,000 died there between 1941 and 1945 due to starvation, overcrowded and unhygienic accommodation and diseases. Another 90,000 were deported to the ghettos and extermination camps in the East, from where only roughly 4,000 returned. Unlike most other ghettos in Nazi occupied East-Central Europe, Terezín was not liquidated at the end of the war. A fraction of its prisoners survived inside of the ghetto walls and were liberated in May 1945. The ghetto has been used for Nazi propaganda purposes and served as destination for old Jewish people from Germany and Austria. In 1944–1945, in an attempt to mislead the world about the genocide of Jews, Terezín was showcased to delegations of the International Committee of the Red Cross.[9]

The Terezín 'Council of Elders', the Jewish 'self-administration' produced a large amount of documents, the lack of paper in the ghetto notwithstanding. However, most of these documents have been destroyed at the end of the war on the orders of the SS Commander of the ghetto. Karel Lágus and Josef Polák describe how all materials relating to the pre-1945 period were taken away and that the search for 'dangerous' documents extended not only to the offices of the 'self-administration', but also to the lodgings of the prisoners. Especially lists and card files of murdered people and of those who were deported to the ghettos and extermination camps in the East were confiscated (cf. Lagus/Polák 1964: 201), as was the physical evidence of death in Terezín: the ashes of people who perished in Terezín were taken away and partly dispersed in the Ohře/Eger river and partly put into an unmarked pit outside of Terezín.

The surviving original documentation was thus either actively used or created after 1st January 1945 (such as for instance a card file of prisoners who mostly survived in the ghetto until its liberation in May 1945), kept illegally by various groups of prisoners, or (re)created after the liberation. Some documents in the central registry or their copies such as the transport lists were hidden and survived its evacuation. A number of prisoners collected and saved various documents: for instance, Karel Herrmann (Heřman) who documented cultural life in the ghetto (cf. Kryl 1986; Štefaniková 2004) or H. G. Adler, one of the future historians of the ghetto. The most extensive set of documents was gathered by Hechalutz, the Zionist youth movement around Zeev Scheck, and was transported to Prague after liberation by his girlfriend Alisah (meanwhile, Scheck was deported to Auschwitz and eventually liberated in Dachau).

Terezín is known for the cultural production of prisoners – however, much of the art work and of prisoners' diaries vanished with their authors following the deportation to extermination camps. Only some manuscripts or pictures were left with friends in Terezín or hidden in the walls or other hide-outs. The Terezín resistance managed to save some important art works that testified about the reality of life in the ghetto, including those of a group of painters employed in the Technical Department who used their access to paper and other material to document the reality of Terezín and who were discovered and deported to the Small Fortress (a concentration camp-like Gestapo prison) in Terezín. Years after the liberation, documents were found in Terezín, for instance the diary of Egon Redlich (1995). On the other hand, in contrast to some of the other major Polish ghettos, very few authentic photos and film footage from the Terezín ghetto are available. The 1942 and 1944 propaganda films and the photos shoot around the filming in 'beautified' Terezín are to be used only with utmost caution (cf. Strusková 2011; Margry 1998). The authentic photo material consists mainly of a series of snapshots taken secretly by a Czech gendarme guarding the ghetto and an extensive series made in Terezín shortly after liberation.

---

9 Of the extensive, though not very recent, literature on Terezín, see mainly Lederer (1983), Adler (1955), Lagus/Polák (1964), Kárný (1991), and the periodicals *Theresienstädter Studien und Dokumente* as well as *Terezínské listy*.

Immediately after liberation, a group of Zionist activists led by Scheck started the Dokumentační akce (or 'Documentation Project'), a Czech (and later Slovak) version of the Jewish documentation initiatives which collected documents and testimonies in many European countries.[10] Within the short period between liberation and immigration to Palestine, the group collected testimonies, documents, photos and artwork documenting the persecution of Jews from Bohemia. In 1946, after Scheck had moved to Palestine, the collection was divided: the larger part was sent to Palestine and a smaller one was placed with the Jewish Museum in Prague. The division of material followed the arrangement submitted to the Jewish Agency, which sponsored the project: duplicate documents including most of the testimonies would go to the JMP and to Palestine, whereas originals would mainly be sent to Palestine. In a clear demonstration that documentation and immigration to Palestine, as well as the struggle for the Jewish state in Palestine, went hand-in-glove for Scheck and his colleagues, the boxes with documents also contained illegal content like weapons. At the same time, documents were collected in the Jewish Museum in Prague where H. G. Adler worked until 1947, when he finally emigrated to the United Kingdom. The collection of the Dokumentační akce and of the JMP was apparently used by the first historiographers of Terezín, Zdeněk Lederer (1983), Karel Lagus and Josef Polák (1964). H. G. Adler (1955) also drew on this material while working on his influential monograph.

---

10 For a comparative history of these projects, see Jockusch (2012).

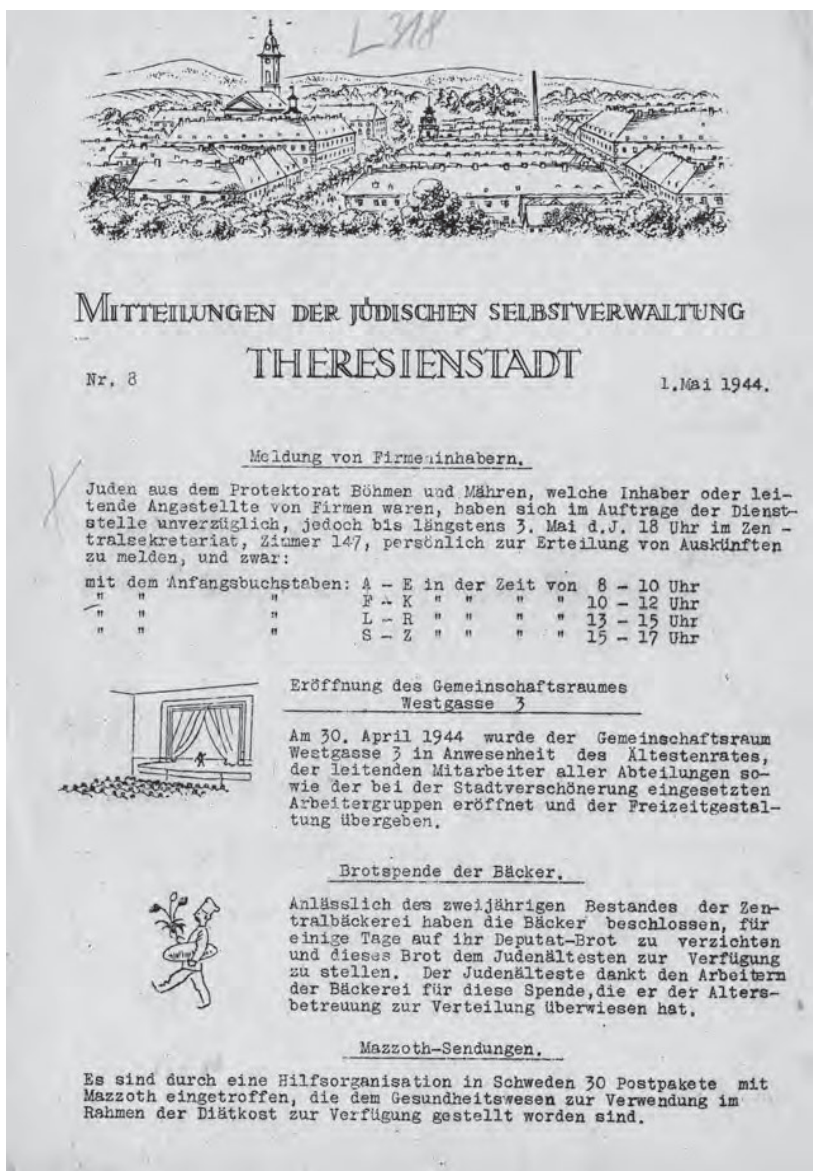

*Figure 2* Daily bulletin of the Jewish self-administration, 1st of May 1944 (Archive of the Jewish Museum in Prague). Originally called orders of the day (Tagesbefehle), the daily bulletins were later renamend and graphically enhanced for the purposes of the Nazi propaganda. The bulletins, in originals or copies, are divided between all major Terezín archives.

As a result, the most important Terezín archives are located in dedicated memorial institutions. Following the Israeli War of Independence, the Documentation Project archive had been stored at the university campus at Mount Scopus, where it was not easily accessible. Only after the 1967 war was the collection moved to Yad Vashem and made available to researchers. Scheck also brought the Hechalutz Terezín collection to Palestine and donated it to the Central Archives for the History of the Jewish People, while Scheck continued to extend it. In 1976, most of the collection was transferred to Yad Vashem, whereas a smaller part was kept in the Central Archives and some materials, especially photos, were moved to Beit Theresienstadt in the Givat Haim kibbutz in Israel. Founded by Terezín survivors (including Zeev and Alisa Scheck) in 1960s, Beit Theresienstadt is a museum, an archive and an educational institution. The Terezín collection in Yad Vashem was later also extended by a collection of transport lists and albums devoted to the activity of various departments of the 'Council of Elders' which was saved by Hermann Weisz and acquired after his death in 1979. A third subcollection contains mostly personal information and documents provided by the former inmates and their families.

The Terezín collection in the Jewish Museum in Prague based on the Documentation Project as well as other materials and over time extended by further acquisitions, was organised later into a form that roughly corresponded to the structure of the Jewish 'self-administration' in Terezín. Therefore, the Terezín collection attempts to partially reconstruct the largely destroyed and fragmented original Terezín documentation. The collection continued to grow since the fall of Communism, especially in conjunction with other projects of the JMP. Some interesting documents were received as part of the oral history project of the museum, another within the public appeal to share documents and photos providing information about their deported and murdered neighbours.

The personal story and the documentation trail of the Dokumentační akce leads also to the history of Beit Terezín. The archive of Beit Theresienstadt houses especially documents donated by the members of the organisation, including important artwork and children newspapers. Very soon after the foundation of the Terezín Memorial in 1947 (originally as Memorial to the Suffering of the Nation), its archive and later 'Documentation Department' was created which collected documents from former inmates and found in Terezín, as well as testimonies of former prisoners. Therefore, any serious researcher of the ghetto has to conduct research at least in these four major

Terezín archives: Beit Terezín, Terezín Memorial, Yad Vashem and the Jewish Museum Prague. Further significant Terezín collections and documents can be found in other archives around the world, for instance in the National Archives in Prague (many of the Terezín related documents were digitised by the Terezín Initiative Institute and are partly accessible online at www.holocaust.cz), the Center for Jewish History in New York (especially in YIVO archives), the Wiener Library in London, NIOD in Amsterdam, the International Tracing Service in Bad Arolsen or the Institute of Contemporary History in Munich. EHRI aims to integrate these resources on Terezín in a dedicated research guide.

## Intellectual Integration: Identification Work and Research Guides

The primary aim of the EHRI Terezín Research Guide is to create a comprehensive, innovative and easy to use guide on the dispersed and fragmented Terezín (Theresienstadt) archival material and to empower further research on the history of the ghetto. The Terezín research guide illustrates the primary raison d'être of EHRI – to connect collections spread in many archives and in more countries. Moreover, EHRI and specifically the research guides it develops demonstrate what a collaborative archival project can achieve and how archivists can redefine their tasks beyond providing physical access and creating finding aids restricted to the local collections. The guide does not aim to make the existing archives irrelevant by placing all information online, but to help researchers identify relevant sources and to connect and compare them to documents in other collections. The guide will function as a gateway to the Terezín archival resources and – as an increasing amount of digitised material appears online (in fact, all four major Terezín archives either have already digitised their collection or are in the process of doing so) – it will point to the respective public online catalogues. We also hope that the guide will catalyse further research in the history of Terezín. While much has already been written and published, there is still no recent synthetic monograph about its history, structure and function, which would integrate Terezín into the broader research on the Holocaust.

In more than one way, the research guides test the challenges that EHRI as a whole needs to meet. The cataloguing standards and data collected from the four major partners show significant differences. While Yad Vashem is a large archive with extensive staff, Beit Terezín is a very small organisation with few staff and limited archival competence. The Terezín Memorial, as a museum funded by the Czech Ministry of Culture, follows museum standards and tends to catalogue documents as individual items. The Beit Terezín archive differentiates between a collection of originals and subject oriented files (which – in turn – often contain copies of the originals). While Yad Vashem only catalogues the material on the file level (with files often containing hundreds of pages), other partner archives provide much more detail information going down to individual documents. Whereas the Terezín finding aid of the Jewish Museum Prague is hierarchical and contains up to ten levels, Yad Vashem's uses subcollections and files, Beit Theresienstadt the file level only, and Terezín Memorial works with separate items. Moreover, the main four Terezín collections integrated in the guide contain a number of copies of items also found in the other.

The archives use different cataloguing systems and standards: whereas Yad Vashem deploys a commercial system with the possibility of standard-compliant export, the Jewish Museum runs an open source cataloguing system with a very flexible metadata schema. The Terezín Memorial stores its data in a simple custom database created without taking into account any set of standards. By the start of the project, Beit Terezín only had short textual descriptions of its files (stored in separate MS Word files). Exporting the information in a compatible format and integrating it into EHRI is no easy task and requires a great deal of mapping and data transformation.

The EHRI research guides team comprises archivists and other experts from the Jewish Museum in Prague, Yad Vashem, Terezín Memorial, and Beit Terezín and has designed a strategy to integrate this diverse material. The team has agreed to keep most of the original finding aids and other descriptive data intact and focus on the structured metadata such as keywords, places, etc. which will make it possible to connect items from different archives. At the very outset of the project, a limited set of principal keywords was defined and a detailed hierarchical list of departments of the 'Council of Elders' (the Jewish 'self-administration' in the ghetto) was created by the Jewish Museum Prague in cooperation with the other team members. Within the project, the partners improved their data and – as much as possible – used the pre-defined metadata.

The Jewish Museum Prague analysed the provided data and where needed created the mappings between the different sets of metadata of the partners. Much attention has been devoted to the geography of the Terezín ghetto. The Jewish Museum Prague created an authoritative list of locations (including

GPS coordinates) inside the ghetto (houses, barracks, crematorium, etc.), which makes it possible to link documents to specific places on the current or historical map of Terezín. In future, the data might be used by the visitors to the site of the former ghetto on their portable devices, as they walk through the town.

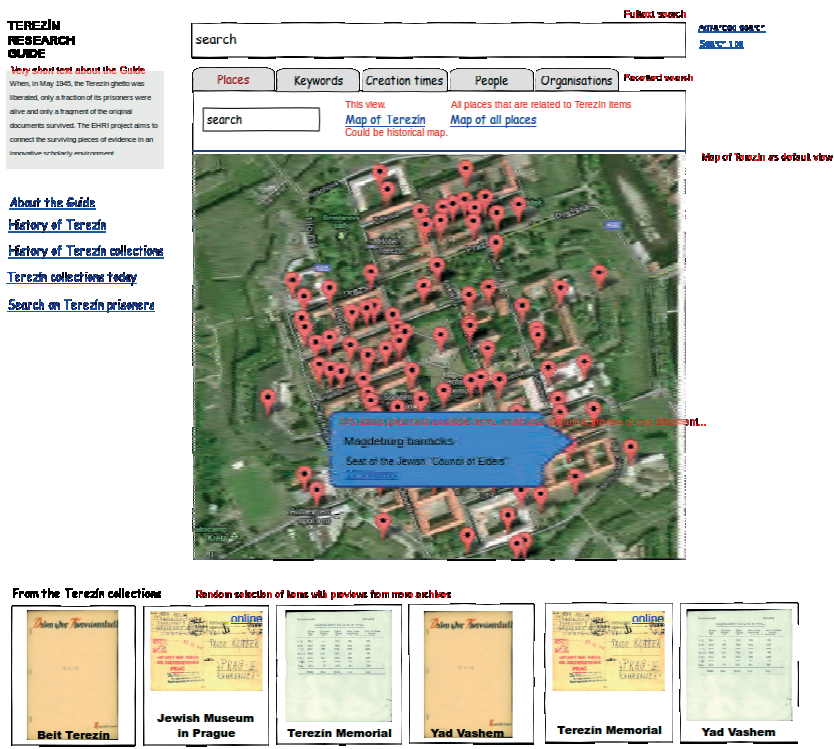

*Figure 3* One of the early designs (wireframes) to approach the documents and files through the map of Terezín

Another principal type of metadata used to connect information and to make the work with the guide productive are databases of Terezín prisoners. The databases can provide a unified authoritative personal reference, as they contain further metadata for contextualisation and/or search. All four archives have access to the database created by the Terezín Initiative Institute in Prague (and in fact, the databases of the Jewish Museum Prague, Beit Terezín and Terezín Initiative Institute have been mapped onto each other), but only the Jewish Museum directly employs this database as personal authority for cataloguing of the archive.

The guide will be designed not only for highly professional users (such as historians), but also with a view to the needs of students, interested members of the public or family members of Terezín prisoners. Therefore, the guides are meant to be used also by people who do not know how to work with traditional archival finding aids and have no extensive historical expertise about Terezín. Short biographic information was prepared for personalities most often referred to, as well as a basic timeline with details about different events and periods in the history of the ghetto. This contextual information will first help users formulate the query itself – for instance by offering information about people, definitions of keywords or descriptions of the functions of the main departments of the Terezín 'self-administration'. On the other end of the search, the retrieved items (files or documents) will be contextualised not only by definitions of metadata, details about places or biographies, but also by placing the document on the chronological scale and making it easy to research related events and documents from the same period. A short history of Terezín, as well as the history of the principal Terezín archival collections, information about the databases of Terezín prisoners and other resources will be available as well.

The guide will differ from the more traditional archival guides or finding aids in that it will avoid ordering information along just one authoritative narrative or path, but will rather provide multiple ways and approaches. In the next year, we will be experimenting with ways which would enable researchers to traverse hierarchies of keywords, through the departments of the 'Council of Elders', and to approach the documents and files through the map of Terezín. The guide will be integrated into the EHRI portal and will share its collaborative features.

## Technical Integration: The Central Point of Access

The aim of the EHRI online portal is to serve as a single point of access to descriptions of Holocaust-related archival sources that are scattered across hundreds of CHIs in and beyond Europe. It will bring together the results of the investigative work carried out elsewhere in the project (and discussed above), provide tools to support discovery and analysis of the integrated information as well as communication and collaboration among Holocaust researchers and archivists. The portal will play a major role in turning currently disparate and fragmented Holocaust documentation into a cohesive corpus; thereby providing an online infrastructure that can enhance existing research, as well as inspire and facilitate new and innovative approaches to the history of the Holocaust.

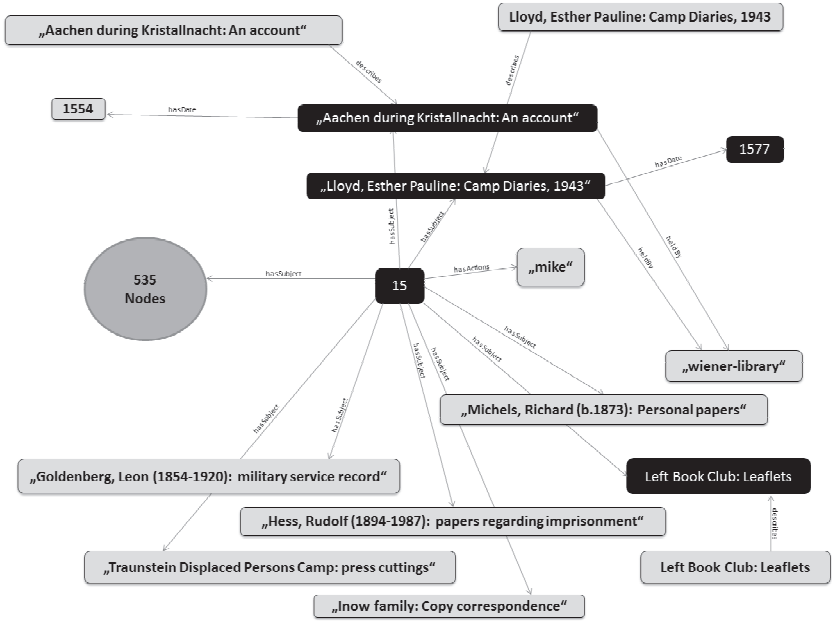

*Figure 4* The EHRI archive graph

The backbone of the EHRI portal is an integration solution to the dispersed archival collections that are relevant to the study of the Holocaust. At the heart of this solution stands the EHRI metadata registry that will be filled with descriptions of relevant archival material harvested from CHIs. One of the main problems we face in this respect is that existing archival metadata is typically expressed in non-standardised and therefore highly heterogeneous forms. We address this problem in various ways such as harmonisation of metadata across sites, but we have also developed a new approach by using a graph database as the basis for our registry. Even though moving away from the tried-and-tested model of relational databases poses a (calculated) development risk, this is in our opinion more than counter-balanced by its benefits. Most importantly, a graph database is flexible with respect to information schemas, and lends itself to the modeling of relationships between data elements in an adaptable and semantically rich fashion. Making best use of such capabilities, our metadata registry will facilitate inter-linkages between integrated descriptions, as well as the embedding of such descriptions in their wider archival and scholarly contexts.

Research users will interact with the information content of the portal via a variety of dedicated tools and services. Taken together, they will form a virtual observatory to enable discovery of, and access to Holocaust related archival sources. Just like astronomers require a (virtual) observatory, Holocaust researchers need a digital infrastructure to study the sources that are until now hidden and often locked away in archives. While we are currently still finalising the formulation of detailed user requirements for the virtual observatory, a few general remarks about some of its features are already possible:

*Multi-lingual Search*

We have shown above how Holocaust-related sources are dispersed across the globe, and how this dispersal requires researchers to negotiate a wide variety of languages, cultures and institutions. To complicate matters further, across countries and institutions, many different languages, perspectives and idioms have been brought to bear when cataloguing and describing these sources. This means that solely bridging across different metadata standards is not a sufficient integration strategy if truly unified access to the archival record of the Holocaust is to become a reality.

The EHRI portal will support unified access by incorporating a multi-lingual terminological database containing a dedicated thesaurus of well-defined subject terms relevant to Holocaust research and a set of authority files. The portal's search interface will use this database to translate and semantically expand user queries. This will enable users to locate all sources that are related to a given subject keyword regardless of differences in regard to provenance, language and/or employed descriptive idioms and terminologies.

*Annotation Service*

Description of archival material can by definition never be final, complete and exhaustive (cf., for instance, Duff/Harris 2002). Even though archives have invested heavily in producing descriptions that are as authoritative and rich as is possible, new research, or a new theoretical paradigm, may always produce a new perspective on a given collection; bring to light connections between collections previously ignored; reveal factual mistakes in existing descriptions and so on. Archival descriptions should therefore not be fixed and monolithic, but 'living' and plural. EHRI will attempt to facilitate both by offering an annotation service.

In the context of the EHRI portal, research users will be enabled to annotate descriptions of archival material in a variety of ways. They may, for instance, use annotations to enhance a given description by attaching a textual note, or by linking it to a term from the EHRI thesaurus, an authority file, or another collection hosted by a different institution. Likewise, annotations could be used to embed an archival collection in its proper scholarly context by relating it to published research outputs such as journal articles, monographs, book chapters and so on. By experimenting with annotations, we are therefore attempting to bridge the gap between researchers and archivists for the benefit of both: researchers are empowered to actively contribute to archival descriptions, while archivists can make use of researchers' expertise and incorporate their suggestions in their own authoritative descriptions. The ultimate outcome should be rich and ever evolving descriptions illuminating the material from many different points-of-views.[11]

*Integrated Helpdesk*

Another important EHRI service to strengthen ties between researchers and archivists is the integrated helpdesk. Such a service is crucially important. Indeed, our work on research user requirements suggests that a lot of Holocaust-related sources are currently found via direct contact between researchers and reference archivists rather than online searches.[12] By offering an integrated helpdesk, we plan to enhance this traditional and popular method of access in the context of the EHRI.

We are currently experimenting with a number of approaches taken from call routing (cf. Lee et al. 1998) and e-mail classifications systems (cf. Youn/McLeod 2007) to establish an automated helpdesk. To this effect, we are building a knowledge base containing semantic representations of archival institutions that hold Holocaust-related collections. In the final implementation, we are planning to give researchers the opportunity to submit their questions as free texts. The helpdesk system will extract relevant information from the texts, evaluate which of the institutions represented in the knowledge base is most suitable to answer, and reply with a list of institutions, ordered by an estimation of their suitability, and including further information that should help researchers to decide which institution to contact to discuss their problems in detail.

---

11 For a good discussion about using annotations and similar technologies to redefine the relationship between the archivist and the scholar and to improve access in an digital environment, see, for instance, Evans (2007), and Yakel/Shaw/Reynolds (2007).

12 That scholarly access to archival material is frequently mediated by a reference archivist has also been stressed by Pugh (1982), Duff/Fox (2006), and Johnson/Duff (2005).

These and a host of further services will provide researchers with a set of useful tools to virtually explore the information content of the EHRI portal, to actively contribute to its development and to find their ways to relevant physical infrastructures. While we are therefore focusing on supporting scholarly research, we are also actively engaged in supporting the work of collection holding institutions. Unlike some of the large aggregators of Holocaust-related archival material which created physical copies of the files and largely separated them from the original archive, EHRI strives to connect researchers to the original institution and its expertise.

Apart from offering institutions the possibility to increase the visibility of their collections through integration of their metadata into the portal, we are currently investigating a number of services that could benefit archives. For instance, even though the portal will not hold digital copies of collections, EHRI may ultimately act as a broker to assist institutions with access to European wide services for data managements and storage. In a similar fashion, we are exploring the possibility of offering small institutions with Holocaust-related collections but no native capabilities to publish their findings aids online with an EHRI hosted archival description service. Finally, we are seeking to offer interested archives a range of Optical Character Recognition and Named Entity Recognition services to support and enhance present and future archival digitisation programmes (cf. Rodriguez et al. 2012).

The EHRI portal concentrates on integrating and enhancing existing capacities via a single point of online access, but by doing so also provides an environment to bring about change in both Holocaust research and archival theory and practice. The portal will offer Holocaust scholars the possibility to find relevant sources across institutional, national and linguistic boundaries and with opportunities to combine, contextualise, exchange and disseminate the results of their research in new ways. It offers CHIs tools and services that increase the online visibility of their collections, improve access to their holdings, and enable innovative methods of engagement with their scholarly readers. The portal is, in other words, a central ingredient in an infrastructure that harnesses digital opportunities to connect people and resources in new ways.

## Conclusion

This article started with a description of the current state of the art in Holocaust research and how it is characterized by many dispersed collections, partly distributed across continents. By connecting collections from different archives and countries with research, EHRI aims to address these important challenges that Holocaust researchers face and make a significant contribution towards more integrated research on the Holocaust. In the first two years of the project, EHRI has worked on the identification of institutions, which hold Holocaust-relevant collections. It has developed a dedicated workflow to survey these collections and has already delivered on national reports and overviews of archives in several European countries. But EHRI does not aim to just describe as many collections as possible but also to dig deeper into the analysis and description of especially challenging cases. A specific effort is dedicated in EHRI to linking sources related to the Terezín ghetto and developing an innovative research guide which will stimulate new research. All the results of the EHRI investigative work including the guide will be made available through a portal as a virtual central point of access. This portal is at the heart of the digital research infrastructure of EHRI.

As discussed, we identified (or we were able to identify) that setting up such a digital research infrastructure will have to overcome a number of problems that are directly connected to different archival cultures. However, we are confident that we have the solutions in place that will deliver a portal, which will connect people and resources in novel ways, with the aim of furthering our understanding of one of the most important events in European history. But a lot of research still needs to be done to find the best possible digital infrastructure for Holocaust sources, which are so dispersed and archived and published to different standards.